\documentclass[aps]{revtex4}
\usepackage{epsfig}
\usepackage{amsmath,amssymb,color}

\parskip=\medskipamount

\newcommand{\re}{\textrm{Re}\,}

\newcommand{\eq}[1]{(\ref{#1})}
\newcommand{\fig}[1]{Fig.\ref{#1}}

\newcommand{\be}{\begin{equation}}
\newcommand{\ee}{\end{equation}}

\newcommand{\barr}{\begin{array}}
\newcommand{\earr}{\end{array}}

\newcommand{\beqn}{\begin{eqnarray}}
\newcommand{\eeqn}{\end{eqnarray}}

\newcommand{\bs}{\begin{subequations}}
\newcommand{\es}{\end{subequations}}

\newcommand\disp{\displaystyle}

\newcommand{\la}{\left<}
\newcommand{\ra}{\right>}

\newcommand{\vep}{\varepsilon}

\begin{document}

\title{Wetting  transition on a one--dimensional disorder}

\author{D.M. Gangardt$^1$, S.K. Nechaev$^2$\footnote{Also at: P.N. Lebedev
Physical Institute of the Russian Academy of Sciences, 119991, Moscow, Russia}}

\affiliation{$^1$School of Physics and Astronomy, University of Birmingham, Edgbaston, Birmingham
B15 2TT, UK \\ $^2$LPTMS, Universit\'e Paris Sud, 91405 Orsay Cedex, France}

\date{\today}

\begin{abstract}
We consider wetting of a one-dimensional random walk on a half--line $x\ge 0$ in a short--ranged
potential located at the origin $x=0$. We demonstrate explicitly how the presence of a quenched
chemical disorder affects the pinning--depinning transition point. For small disorders we develop a
perturbative technique which enables us to compute explicitly the averaged temperature (energy) of
the pinning transition. For strong disorder we compute the transition point both numerically and
using the renormalization group approach. Our consideration is based on the following idea: the
random potential can be viewed as a periodic potential with the period $n$ in the limit
$n\to\infty$. The advantage of our approach stems from the ability to integrate exactly over
all spatial degrees of freedoms in the model and to reduce the initial problem to the analysis of
eigenvalues and eigenfunctions of some special non-Hermitian random matrix with disorder--dependent
diagonal and constant off-diagonal coefficients. We show that  even for strong disorder the
shift of the averaged pinning point of the random walk in the ensemble of random realizations of
substrate disorder is indistinguishable from the pinning point of the system with preaveraged (i.e.
annealed) Boltzmann weight.
\end{abstract}

\maketitle

\section{Introduction}
\label{sect:1}

Wetting is one of the most intensively studied phenomena of statistical physics of interfaces. In a
very general setting wetting implies the interface pinning by a solid impenetrable substrate.
Problems of interface statistics in the presence of a hard wall were addressed in many publications
(see, for example, \cite{abraham} and references therein). The most interesting questing concerns
the nature of the wetting or pinning-depinning transition of the interface controlled by parameters
of its interactions with the substrate. Here we study the case when the substrate is inhomogeneous,
so the wetting transition occurs in the presence of disorder.

The pinning--depinning transition in models of wetting in presence of quenched disorder was studied
by many research groups since the middle of 80th. In 1986 Forgacs {\em et al} \cite{forgacs}
developed a perturbative renormalization group approach to the (1+1)--dimensional wetting subject
to a disordered potential along the substrate. Around the same time Grosberg and Schakhnovich
\cite{gro}  applied the RG technique for studying an equivalent problem of the localization
transition in ideal heteropolymer chains with quenched random chemical (primary) structure at a
point--like potential well in a D--dimensional space. Many conclusions of \cite{gro} for D=3 agree
with those of \cite{forgacs}. Both approaches provide important information about the
thermodynamics near the point of transition from  delocalized (depinned) to localized (pinned)
regimes in the presence of quenched chemical disorder.

However some crucial questions of pinning--depinning transition in a quenched random potential
still remains  open. One of the most intriguing problems is the determination of the averaged
transition temperature, $T_{\rm q}$, for quenched chemical disorder. Since the temperature enters
into problem through the Boltzmann weight $\beta = e^{u_m/T}$, where $u_m$ is the energy of $m$-th
interface segment, one may attempt to relate the transition point to the temperature $T_{\rm a}$
for annealed chemical disorder with preaveraged Boltzmann weight, $\la\beta\ra=\la e^{u_m/T}\ra$.
The RG approaches \cite{forgacs,gro} claim $T_{\rm q}=T_{\rm a}$ in the thermodynamic limit. In
1992 Derrida, Hakim and Vannimenus \cite{derrida} have reconsidered the (1+1)--dimensional model of
wetting and have shown by a different  RG technique that the disorder is marginally relevant, i.e.
any infinitely small disorder displaces the averaged transition point $T_{\rm q}$ in the ensemble
of quenched sequences from the transition point $T_{\rm a}$ in ensemble of sequences with initially
preaveraged (annealed) Boltzmann weight. Subsequently other works \cite{stepanow,tang}  arrived at
the same conclusion. The equivalent problem of localization transition of a random walk has been
also deeply studied in mathematical literature. In \cite{alexander} it was rigorously proven that
in systems with  return probability which scales as $\sim N^{-\alpha}$, where $N$ is the number of
steps,  the phase transition curves for quenched and annealed systems coincide for $1<\alpha<3/2$
for small disorder and are different for $3/2<\alpha<2$ though they are very close numerically
(again for small disorder). The similar conclusion has been drawn in the work \cite{toninelli} by
an alternative method. In other work \cite{giacomin} it has been shown rigorously that the disorder
is marginally relevant for $\alpha >3/2$. As for the case $\alpha = 3/2$ there is no definite
answer (even for small disorder)  whether the results for  quenched and annealed disorder coincide.
The value  $\alpha=3/2$ of the critical exponent considered here is, therefore, of particular
interest.

In our paper we  demonstrate explicitly how the presence of quenched chemical disorder affects the
pinning--depinning transition point. For small disorder we develop a perturbation theory which
enables us to compute explicitly the transition temperature  of the system. The advantage of our
approach, which borrows the basic idea from \cite{nech_nai}, stems from the ability to integrate
exactly over all $N$ spatial degrees of freedom  of the model and to reduce the initial problem to
the analysis of eigenvalues and eigenfunctions of some special non--Hermitian $N\times N$ random
matrix with disorder--dependent diagonal and constant off-diagonal elements. Our approach is based
on the following general idea: the random potential can be viewed as a periodic potential with the
period $N$ in the limit $N\to\infty$.

The paper is organized as follows. In Section \ref{sect:2} we describe the model and derive basic
equations. In Section \ref{sect:4} we develop the perturbation theory for eigenvalues of our random
matrix and derive the corresponding expressions for the averaged temperature of pinning--depinning
transition for any type of disorder (not necessary to be Gaussian). The simple renormalization
group approach is developed in Section \ref{sect:rg}, while the numerical analysis of general
analytic equations is performed in Section \ref{sect:3}. In Conclusion we summarize our
results and pose some  new questions.

\section{Random matrix formulation of wetting problem}
\label{sect:2}

The problem of fluctuating interface in thermodynamic equilibrium maps into an equivalent problem
one-dimensional random walk  on a half--line $x_m\ge 0$, where $m$ is the discrete time. On the
line $x=0$  additional  Boltzmann weights $\beta_m=e^{u_m/T}$ account for the potential interaction
of the fluctuating interface (the random walk) and  the impenetrable substrate. As soon as the
running time $t=m$  can be associated with the current coordinate along
the wall, one can say that $u_m$ is the interaction energy of an interface with a substrate at a
position $m$. If the interaction energies $u_m$ are arbitrary, then the  set $\{u_m\}=\{u_1,u_2,
..., u_N\}$ represents the quenched random interaction of the interface with the substrate.

\begin{figure}[ht]
\epsfig{file=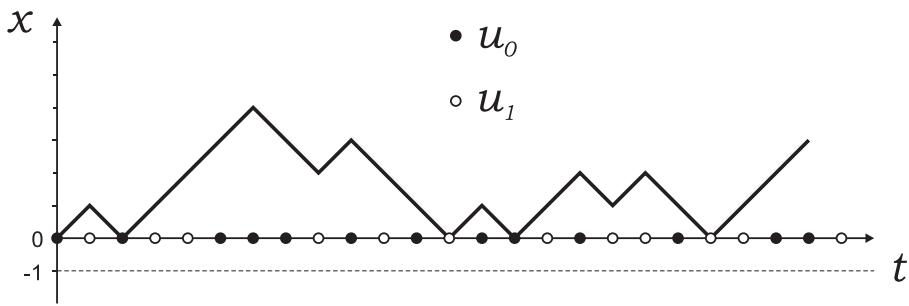,width=12cm} \caption{Wetting in a random potential on a line.}
\label{fig:1}
\end{figure}

Consider a random walk in a semi-axis $x > 0$ which represents the height of the fluctuating
interface interacting randomly with a surface situated at $x=1$. The probability  of the random
walk interacting with random surface potential $\{u_m\}$ to be found at the position $x$ after $N$
steps will be denoted by $G_N (x)$.  This function   satisfies the following recursion relation
\be
\left\{\barr{ll} G_{N+1}(x) = \frac{1}{2}G_N(x-1) + \frac{1}{2} G_N(x+1) +
  \frac{1}{2}(\beta_N-1)
\delta_{x,1} G_N(x+1) &
\qquad x \ge 1 \medskip \\
G_N(x) = 0 & \qquad x=0 \medskip \\
G_{N=0}(x) = \delta_{x,1}
\end{array} \right.
\label{eq:6}
\ee
In the presence of the disordered potential $u_m \neq 0$ the equation (\ref{eq:6}) does not
conserve normalization of the propagator $G_N (x)$ so it has to be explicitly normalized  after $N$
steps. The typical configuration of the random surface is depicted in \fig{fig:1} for bimodal
disorder $\{u_m\}= u_0$ or $u_1$. Let us stress however that our considerations are quite general
and are not restricted to any specific type of a substrate disorder.
To answer the question about the location of the pinning transition find it is more convenient to
change distribution of  $\beta_m$ by changing parameters of $u_m$ at fixed temperature $T$ which is
conventionally set equal to unity for the rest of the article.

Now we define explicitly what is the pinning (or localization) of the random interface in
(1+1)--dimensional wetting problem. Consider the mean--square end-to-end distance, $\la x^2(N) \ra$
of the random interface of length $N$
\be
\la x^2(N) \ra = \frac{\disp \sum_{x=0}^{\infty} x^2 G_N(x)}{\disp \sum_{x=0}^{\infty} G_N(x)}
\label{eq:7a}
\ee
There exists some critical value of the energy, $u_{\rm tr}=\ln \beta_{\rm tr}$, which separates
two different types of behavior of the mean--square end-to-end distance in the thermodynamic limit:
\be
\la x^2(N) \ra\Big|_{N\gg 1} =\begin{cases} \Lambda_1(u) & \mbox{for $u>u_{\rm tr}$} \\ \Lambda_2
(u) N & \mbox{for $u<u_{\rm tr}$}
\end{cases}
\label{eq:7}
\ee
where $\Lambda_{1,2}(u)$ are some positive energy--dependent constants independent of $N$. The
value $u_{\rm tr}$ is called the energy of the pinning transition. Formally speaking, for all
$u>u_{\rm tr}$ the Eq.\eq{eq:6} has a discrete spectrum, and for $u<u_{\rm tr}$ Eq.\eq{eq:6} has a
continuous one. Slightly above the transition point the following critical behavior of the free
energy $F_N (u) =\ln G_N$ of an $N$--step random walk in a half--space $x\ge 0$ is expected:
\be
f(u)=\lim_{N\to\infty} \frac{1}{N}F_N(u)\equiv \lim_{N\to\infty} \frac{1}{N} \ln
G_N\Big|_{u \to u^+_{\rm tr}} = {\rm const}\; (u-u_{\rm tr})^{\alpha}
\label{eq:8}
\ee
where $\alpha$ is the critical exponent defining the order of the phase transition. In the absence
of disorder it was shown  \cite{no_disorder} that $\alpha=2$ which  corresponds to an ordinary 2nd
order phase transition. In the next subsection we reproduce these calculations. The reason for doing
this is to define our notation and introduce important concepts which we shall use in
the more complex disordered case.

\subsection{Wetting in absence of substrate disorder}
\label{sect:2:1}

We review briefly the situation with no chemical disorder , \textit{i.e}. when all interaction
energies take the same value, $\{u_m\}=\{u,u,\ldots,u\}$. Hence, all the Boltzmann weights in
Eq.\eq{eq:6} are equal, i.e. $\beta_m=\beta$ for $m=1,...N$.

Using the discrete $\sin$--Fourier transform and introducing the generating function, we define the
function $G(q,s)$ as follows
\be
G(q,s) = \sum_{x=0}^{\infty} \sin q x \sum_{N=0}^{\infty} s^N G_N(x)
\label{eq:9}
\ee
This function satisfies the following integral equation
\be
\frac{1}{s} G(q,s) - \frac{\sin q}{s} = \cos q\, G(q,s) + \frac{1}{\pi}(\beta-1) \sin q
\int_{0}^{\pi} G(q',s) \sin 2q' dq' .
\label{eq:10}
\ee
Introducing
\be
A(s)= \int_{0}^{\pi} G(q,s) \sin 2q\, dq
\label{eq:11}
\ee
we can rewrite \eq{eq:10} as an algebraic equation for $A(s)$:
\be
A(s) = \int_0^{\pi} \frac{\sin q \sin 2q}{1-s \cos q} dq + \frac{s}{\pi}(\beta-1)A(s)
\int_0^{\pi}\frac{\sin q \sin 2q}{1-s \cos q} dq
\label{eq:12}
\ee
Solving it we get
\be
A(s) = \frac{1}{\disp \left[\int_0^{\pi} \frac{\sin q \sin 2q}{1-s \cos q} dq\right]^{-1} -
\frac{s}{\pi}(\beta-1)}=\frac{\pi}{s}\, \frac{(1-\sqrt{1-s^2})^2}{s^2-(\beta-1)(1-\sqrt{1-s^2})^2}
\label{eq:13}
\ee
which  allows us to write the complete expression for the function $G(q,s)$:
\be
G(q,s) = \frac{\sin q}{1-s \cos q}\left(1+\frac{s(\beta-1)}{\pi}A(s)\right)
\label{eq:14}
\ee
The inverse Fourier transform applied to \eq{eq:14} gives us
\be
G(x,s)=\frac{2}{\pi}\int_0^{\pi}G(q,s) \sin q x\, dq = 2\left(\frac{1}{s}+
\frac{\beta-1}{\pi}A(s)\right) \left(\frac{1-\sqrt{1-s^2}}{s}\right)^x .
\label{eq:15}
\ee
Performing the summation  over all $x\ge 0$ we arrive at the following expressions for
the functions $G(s)=\sum_{x=0}^{\infty} G(x,s)$ and $\sum_{x=0}^{\infty} x^2 G(x,s)$:
\be
\barr{l} \disp G(s)=\sum_{x=0}^{\infty} G(x,s) =
2\left(1+\frac{s(\beta-1)}{\pi}A(s)\right)
\frac{1}{s-1+\sqrt{1-s^2}} \\
\disp \sum_{x=0}^{\infty} x^2 G(x,s) = \disp 2\left(1+\frac{s(\beta-1)}{\pi}A(s)\right) \frac{(s^2
+ (2+s)(-1+\sqrt{1-s^2}))}{(s-1+\sqrt{1-s^2})^3} \earr
\label{eq:16}
\ee
The first equation in \eqref{eq:16} is the basis for the derivation of the free energy
$f=\lim_{N\to\infty}\ln G_N$ and we have
\be
G_N = \frac{1}{2\pi i} \oint G(s) s^{-N-1}\, ds ,
\label{eq:17}
\ee
while the second equation in \eqref{eq:16} provides the mean--square end-to-end distance
\be
\la x^2 (N) \ra = G_N^{-1}\, \frac{1}{2\pi i} \oint \, s^{-N-1}\, ds \sum_{x=0}^{\infty} x^2 G(x,s) .
\label{eq:17a}
\ee
The integrals in Eqs.~(\ref{eq:17},\ref{eq:17a}) are performed along a contour in the complex plane
of $s$ which lies inside the unit circle. The function $G(s)$  is analytic in the whole complex
plane except for the branch cuts on the real axis for  $|\re s|>1 $. In the localized phase the
function $G(s)$ has poles for $|s| < 1$ on the real axis. In the thermodynamic limit $N\to\infty$
the pinning transition is indicated by appearance of the pole  at $s\to s_{\rm tr}=1$, i.e. is
determined by the point of divergence of the function $G(s)$ for $s=1$, which, in turn, diverges
when the denominator of $A(1)$ tends to zero as $\beta$ approaches the transition point $\beta_{tr}
=  e^{u_\textrm{tr}}$. Thus, we have the following equation for $\beta_{\rm tr}$ (see
Eq.\eq{eq:13})
\be
s_{\rm tr} - \sqrt{\beta_{\rm tr}-1}\left(1-\sqrt{1-s_{\rm tr}^2}\right)\bigg|_{s_{\rm tr}=1}=0;
\qquad \beta_{\rm tr} \equiv e^{u_{\rm tr}} = 2; \qquad u_{\rm tr}=\ln 2\; .
\label{eq:18}
\ee

In the localized phase the free energy $f$ in the thermodynamic limit $N\to\infty$  is dominated by
the closest to zero pole $s_0(\beta)$  of the function  $G(s)$ in \eq{eq:16} for some fixed value
of $\beta$:
\be
s_0-\sqrt{\beta-1}\left(1-\sqrt{1-s_0^2}\right)=0; \qquad
s_0=\frac{2\sqrt{\beta-1}}{\beta} .
\label{eq:19}
\ee
Calculating the contribution of this pole to the integral in
Eq.~(\ref{eq:17}), so that:
\be
G_N = \lim_{N\to\infty} \frac{1}{2\pi i} \oint G(s) s^{-N-1} ds = s_0^{-N} =
\left(\frac{2\sqrt{\beta-1}}{\beta}\right)^{-N}
\label{eq:20}
\ee
we arrive at desired expression of the free energy $f$ in the thermodynamic limit
\be
f = - \ln s_0 = \ln \frac{2\sqrt{\beta-1}}{\beta} .
\label{eq:21}
\ee
In the vicinity of the transition point we  write $\beta = \beta_{\rm tr} + \delta$,
where $\delta\ll\beta_{\rm tr}$. Expanding \eq{eq:21} near $\beta_{\rm tr}=2$,
we get
\be
f(\beta \to \beta_{\rm tr}) = \frac{\delta^2}{8} = \frac{(\beta-\beta_{\rm tr})^2}{8} =
\frac{(u-u_{\rm tr})^2}{2}
\label{eq:23}
\ee
Comparing \eq{eq:23} to \eq{eq:8} we conclude that $\alpha =2$ and hence the pinning transition on
a homogeneous substrate is the standard 2nd order phase transition in accordance with the results
\cite{no_disorder}.

\subsection{Wetting in a periodic bimodal potential}
\label{sect:2:2}

Consider now the wetting in a substrate potential with a {\em bimodal} periodic distribution  of
energies $\{u\}= \{u_0,u_1,u_0,u_1,...,u_0,u_1\}$.  We denote by $u_0$ and $u_1$ the energies
belonging to  the even/odd time slices  correspondingly. This problem has been addressed for the
first time in \cite{nech_zhang} and then considered in much more general setting in subsequent
publications \cite{swain,burk,bauer,mont}. The reason to reconsider this problem is basically
methodological: we solve this problem in a matrix form and then in Section \ref{sect:2:3}
generalize this matrix approach to a substrate with arbitrary period of disorder.

The master equation for the function $G_N(x)$ which generalizes \eq{eq:6} is as follows:
\be
\left\{\barr{l} G_{2N+1}(x) = \frac{1}{2} G_{2N}(x-1) + \frac{1}{2} G_{2N}(x+1)+ \frac{1}{2}
(\beta_{0}-1) \delta_{x,1} G_{2N}(x+1) \medskip \\ G_{2N+2}(x) = \frac{1}{2} G_{2N+1}(x-1) +
\frac{1}{2} G_{2N+1}(x+1)+ \frac{1}{2} (\beta_{1}-1) \delta_{x,1} G_{2N+1}(x+1) \medskip \\
G_N(x=0) = 0 \medskip \\
G_{N=0}(x) = \delta_{x,1} \earr \right. ,
\label{eq:24}
\ee
where $\beta_{0,1} = e^{u_{0,1}}$ are the corresponding Boltzmann weights.

Define odd and even functions $G_N$:
\be
\begin{cases} G_{2N}(x) = W_N(x) \medskip \\ G_{2N+1}(x) = V_N(x) \end{cases}
\ee
Rewrite \eq{eq:24} in Fourier space using functions $W_N(x)$ and $V_N(x)$:
\be
\left\{\barr{l} \disp V_N(q) = \cos q\, W_N(q) + \frac{\sin q}{\pi}(\beta_0-1) \int_0^{\pi} W_N(q')
\sin 2q'\, dq'
\medskip \\ \disp  W_{N+1}(q) = \cos q\, V_N(q) +
\frac{\sin q}{\pi}(\beta_1 -1) \frac{2}{\pi} \int_0^{\pi} V_N(q') \sin 2q'\, dq' \medskip \\
W_{N=0}(q) = \sin q \earr \right.
\label{eq:25}
\ee
Introducing the generating functions $W(q,s)$ and $V(q,s)$
\be
W(q,s)=\sum_{N=0}^{\infty} W(q) s^N; \qquad V(q,s)=\sum_{N=0}^{\infty} V(q) s^N
\label{eq:26}
\ee
we can write a closed system of integral equations
\be
\left\{\barr{l} \disp V(q,s) = \cos q\,  W(q,s) + \frac{\beta_0 -1}{\pi} \sin q \int_0^{\pi} W(q',s)
\sin 2q' dq'  \medskip \\ \disp W(q,s) = \sin q + s \cos q\, V(q,s) + s \frac{\beta_1 -1}{\pi} \sin q
\int_0^{\pi} V(q',s) \sin 2q' dq' \earr \right. .
\label{eq:27}
\ee
It is worth noting that the variable $s$ plays the role of the fugacity of the two-step block and
is no longer associated with a single step as, for example, in Eq.\eq{eq:6}.

Equations \eq{eq:27} allow for a very convenient matrix formulation which could be later easily
generalized to longer periods. Introduce the matrices
\be
\hat A = \left(\begin{array}{cc}  0 & 1 \medskip \\ 1 & 0 \end{array}\right); \quad \hat B =
\left(\begin{array}{cc} \beta_0 -1 & 0 \medskip \\ 0 & \beta_1 -1 \end{array}\right); \quad \hat
M_s = \left(\begin{array}{cc} 0 & 1 \medskip \\ s & 0 \end{array}\right); \quad \hat I =
\left(\begin{array}{cc}  1 & 0 \medskip \\ 0 & 1 \end{array}\right)
\label{eq:28}
\ee
and vectors
\be {\bf G}(q,s) = \left(\begin{array}{c} V(q,s) \medskip \\ W(q,s) \end{array}\right); \quad
{\bf F}(q) = \left(\begin{array}{c} 0 \medskip \\ \sin q \end{array}\right)
\label{eq:28a}
\ee
Rewriting Eq.\eq{eq:27} using \eq{eq:28}--\eq{eq:29}, we obtain  equation for the vector
function ${\bf G}(q, s)$
\be
{\bf G}(q,s) = {\bf F}(q) + \cos q\, \hat M_s\, {\bf G}(q,s) + \frac{\sin q}{\pi} \hat B \hat M_s
\int_0^{\pi} \, {\bf G}(q',s) \sin 2q'\, dq'
\label{eq:29}
\ee
For further analysis it is convenient to rewrite \eq{eq:29} in the following form
\be
{\bf G}(q,s) = \left(\hat I - \cos q\, \hat M_s\right)^{-1} {\bf F}(q) + \frac{\sin q}{\pi}
\left(\hat I - \cos q\, \hat M_s\right)^{-1} \hat B \hat M_s \int_0^{\pi} {\bf G}(q',s) \sin 2q'\, dq'
\label{eq:30}
\ee
Define now
\be
{\bf Q}(s) = \int_0^{\pi} {\bf G}(q,s) \sin 2q \, dq
\label{eq:31}
\ee
(compare to Eq.\eq{eq:11}). The solution for ${\bf Q}(s)$ reads
\be
{\bf Q}(s) = \left[\hat I - \frac{1}{\pi} \int_0^{\pi} dq\, \sin q \sin 2q\,\left(\hat I - \cos q
\, \hat M_s \right)^{-1} \hat B \hat M_s \right]^{-1} \int_0^{\pi} \left(\hat I - \cos q\, \hat M_s
\right)^{-1} {\bf F}(q) \sin 2q\, dq
\label{eq:32}
\ee
This equation  extends the solution \eq{eq:13} to the periodic bimodal potential $\{u\}=
\{u_0,u_1,u_0,u_1,...,u_0,u_1\}$. Setting parameter $s$ to its critical value $s=s_{\rm tr}=1$ and
calculating explicitly the integrals we arrive at the following result
\begin{equation}
  \label{eq:q_explicit}
 \mathbf{Q}(s) = \left(
   \begin{array}{cc}
     \frac{1}{2-\beta_1} & 0 \\
     0 & \frac{1}{2-\beta_0}
   \end{array}\right)
 \left(
   \begin{array}{c}
     2\pi \\ 0
   \end{array}\right)
\end{equation}
implying the phase  transition at $\beta_1=2$. It is clear from the above expression that the value
of $\beta_0$ is actually \textit{irrelevant}. This fact reflects the peculiarity of the microscopic
model: after an even number of steps the random walk has exactly zero probability to reach $x=2$ from
which transition to $x=1$ is controlled by $\beta_0$. Mathematically it is reflected in the
orthogonality of $\mathbf{F}(q)$ to the eigenvector belonging to the eigenvalue $1/(2-\beta_0)$ in
Eq. (\ref{eq:q_explicit}). We shall see in the next Subsection that this peculiarity persists for
arbitrary even periods of the substrate potential.

\subsection{Wetting in a potential with arbitrary period
length}
\label{sect:2:3}

We can straightforwardly generalize the approach developed in the previous Section to the case of a
substrate potential with the period $n$:
\be
\{u\}=\{\ldots,\overbrace{u_0,u_1,...,u_{n-1}}^{\rm period},...,\overbrace{u_0,u_1,...,u_{n-1}}^{\rm
period},\ldots\}
\label{eq:periodic}
\ee
i.e. the total substrate consists of $\ell=N/n$ copies of random subchains of length $n$ each. The
equations (written already in the Fourier space) which extend \eq{eq:25} to the case of repeating
$n$--periodic potential are as follows
\be\left\{
\barr{l} \disp G_N^{(1)}(q) = \cos q\, G_N^{(0)}(q) + \frac{\sin q}{\pi}(\beta_0-1) \int_0^{\pi}
G_N^{(0)}(q') \sin 2q'\, dq' \medskip \\
\disp G_N^{(2)}(q) = \cos q\, G_N^{(1)}(q) + \frac{\sin q}{\pi}(\beta_1-1) \int_0^{\pi}
G_N^{(1)}(q') \sin 2q'\, dq' \medskip \\
... \medskip \\
\disp G_N^{(n-1)}(q) = \cos q\, G_N^{(n-2)}(q) + \frac{\sin q}{\pi}(\beta_{n-2}-1) \int_0^{\pi}
G_N^{(n-2)}(q') \sin 2q'\, dq' \medskip \\
\disp G_{N+1}^{(0)}(q) = \cos q\, G_N^{(n-1)}(q) + \frac{\sin q}{\pi}(\beta_{n-1}-1) \int_0^{\pi}
G_N^{(n-1)}(q') \sin 2q'\, dq' \medskip \\
\disp G_{N=0}^{(1)}(q) = \sin q \earr \right.
\label{eq:35}
\ee
As in the case of bimodal disorder rewrite Eqs. \eq{eq:35} in a matrix form
\be
{\bf G}(q,s) = {\bf F}(q) + \cos q\, \hat M_s\, {\bf G}(q,s)+  \frac{\sin
  q}{\pi} \hat B \hat M_s\int_0^{\pi}
 {\bf G}(q',s) \sin 2q'\, dq'
\label{eq:36}
\ee
where
\be
\hat B = \left(\begin{array}{cccccc}
  \beta_0-1 & 0 & 0 & \dots & 0 & 0 \\
  0 & \beta_1-1 & 0 &   & 0 & 0 \\
  0 & 0 & \beta_2-1 &   & 0 & 0 \\
  \vdots &   &   & \ddots &   & \vdots \\
  0 & 0 & 0 &   & \beta_{n-2}-1 & 0 \\
  0 & 0 & 0 & \dots & 0 & \beta_{n-1}-1
\end{array} \right),\qquad
\hat M_s = \left(\begin{array}{cccccc}
  0 & 0 & \dots & 0 & 0 & 1 \\
  1 & 0 &   & 0 & 0 & 0 \\
  0 & 1 &   & 0 & 0 & 0 \\
  \vdots &  &  \ddots &  &   & \vdots \\
  0 & 0 &   & 1 & 0 & 0 \\
  0 & 0 & \dots & 0 & s & 0
\end{array} \right),
\label{eq:37}
\ee
and
\be {\bf G}(q,s) = \left(\begin{array}{l} G_1(q,s) \medskip \\ \vdots \medskip \\ G_{n-1}(q,s) \medskip
\\ G_n(q,s) \end{array}\right);
\quad {\bf F}(q) = \left(\begin{array}{c} 0 \medskip \\ \vdots \medskip \\ 0 \medskip \\ \sin q
\end{array}\right)
\label{eq:37a}
\ee
Introducing (as in Eq.\eq{eq:31})
\be
{\bf Q}(s) = \int_0^{\pi} {\bf G}(q,s) \sin 2q \, dq
\label{eq:38}
\ee
we get for \eq{eq:36}
\be
\mathbf{Q} (s)  - \frac{1}{\pi} \int_0^{\pi} dq \, \sin q \sin 2q\, \left[\hat I - \cos
  q\, \hat M_s \right]^{-1}\, \hat B \hat M_s   {\bf Q}(s)
= \int_{0}^{\pi} dq\, \sin 2q\left[\hat I - \cos q\, \hat M_s
\right]^{-1} {\bf F} (q).
\label{eq:39}
\ee
where by $\hat I$ we have denoted the $N\times N$ unit matrix.

\begin{figure}[ht]
\epsfig{file=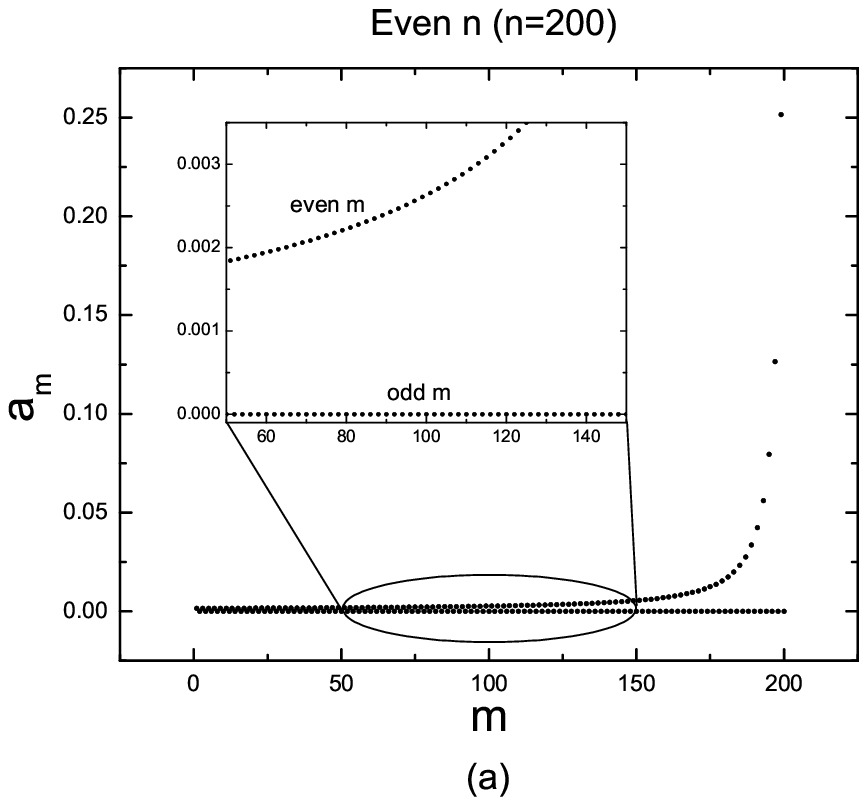,width=8cm} \epsfig{file=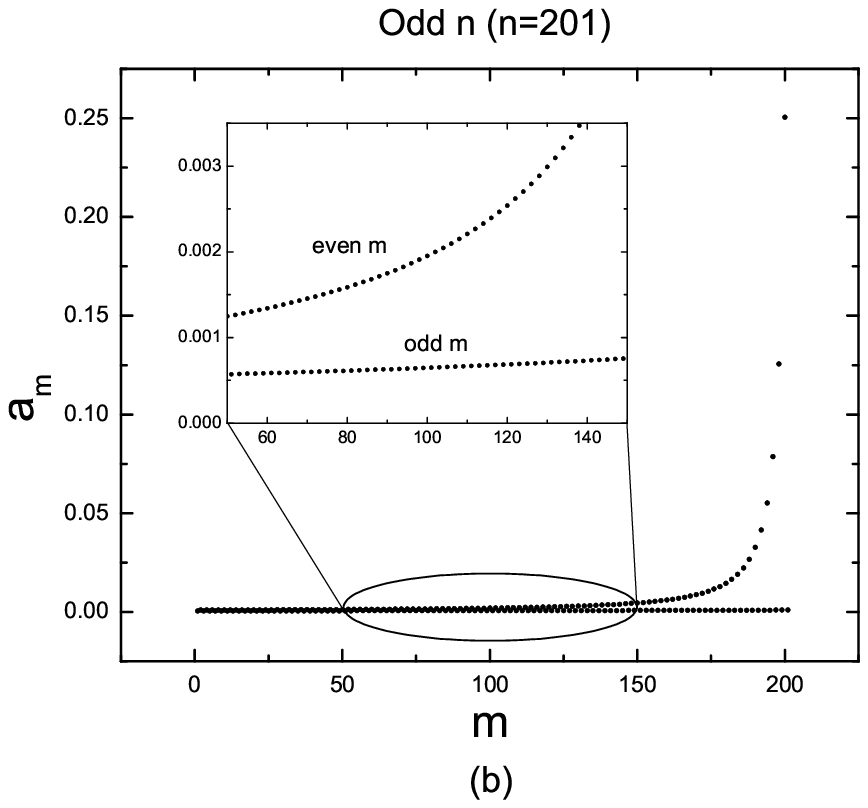,width=8cm} \caption{Behavior of the
coefficients $a_m$ for even (a) and odd (b) values of total length $n$.}
\label{fig:am}
\end{figure}

It can be checked that $\hat B \hat M_s = \hat M_s \hat B'$,  where
\be
\hat B' = \left(\begin{array}{cccccc}
  \beta_1-1 & 0 & 0 & \dots & 0 & 0 \\
  0 & \beta_2-1 & 0 &   & 0 & 0 \\
  0 & 0 & \beta_3-1 &   & 0 & 0 \\
  \vdots &  &   & \ddots &   & \vdots \\
  0 & 0 & 0 &   & \beta_{n-1}-1 & 0 \\
  0 & 0 & 0 & \dots & 0 & \beta_{n}-1
\end{array} \right), \qquad
\beta_n=\beta_0 .
\label{eq:40}
\ee
Introducing the modified vector ${\bf R}(s)=\hat B'\, {\bf Q}(s)$ we rewrite  Eq.
(\ref{eq:q_explicit}) in the following form:
\be
 \hat T_s \mathbf{R} (s) =\left[ \hat {B'}^{-1} - \frac{1}{\pi} \int_{0}^{\pi} dq \, \sin q \sin 2q\,
\left[\hat I - \cos  q\, \hat M_s \right]^{-1}\, \hat M_s  \right]{\bf R}(s) = {\bf C}(s)
\label{eq:42}
\ee
 with the right hand side
\be
{\bf C}(s)=\int_{0}^{\pi} dq\, \sin 2q\left[\hat I - \cos q\, \hat M_s  \right]^{-1} {\bf F}(q) .
\label{eq:42a}
\ee
The matrix $T_s$ in Eq. (\ref{eq:42}) can be explicitly written as
\be
\hat T_s \equiv \left(\begin{array}{cccccc}
  a_0-\dfrac{1}{\beta_1-1} & a_1 & a_2 & \dots & a_{n-2} & a_{n-1} \\
  a_{n-1} & a_0-\dfrac{1}{\beta_2-1} & a_1 &  & a_{n-3} & a_{n-2} \\
  a_{n-2} & a_{n-1} & a_0-\dfrac{1}{\beta_3-1} & & a_{n-4} & a_{n-3} \\
  \vdots &  &  & \ddots &  & \vdots \\
  a_2 & a_3 & a_4 &   & a_0-\dfrac{1}{\beta_{n-1}-1} & a_1 \\
  a_1 & a_2 & a_3 & \dots & a_{n-1} & a_0-\dfrac{1}{\beta_{n}-1}
\end{array} \right)
\label{eq:45}
\ee
and has disorder on its main diagonal only. The nonrandom elements $a_m$ are given by the following
integrals
\be
a_m \equiv a_m(s) = \frac{1}{\pi} \int_0^{\pi} \frac{s \sin q\, \sin 2q\, \cos^{n-m-1} q}{1-s
\cos^n q}\,dq \label{eq:46} .
\ee
For $n$ even the symmetry of the integrand implies that $a_m$ are zero for $m$ odd.  The structure
of the matrix $T_s$ is therefore similar to that of Eq. (\ref{eq:q_explicit}): it does not mix
vectors having all but odd/even zero elements. In the following we call these subspaces odd and
even sectors. It is important to note that $\mathbf{C}(s)$ belongs to the odd sector and therefore
is affected by $\beta_m$ with $m$ odd only.  For $n$ odd the integrand in the Eq.(\ref{eq:46}) has
no  symmetric properties and all elements $a_m$ are nonzero. In this case odd and even sectors are
mixed by the matrix $T_s$ and the physical behavior of the polymer depends on all the values of the
disorder potential.  The behavior of the coefficients $a_m$ is depicted in \fig{fig:am} separately
for even and odd total lengths, $n$. In what follows we restrict ourselves to the case of odd
values of $n$.  This avoids the discussion of the somewhat pathological situation where the pinning
transition is insensitive to a macroscopic number of values of the disorder. This situation
arises due to the impossibility of return to the origin after an odd number of steps of our random
walk and can be eliminated by a different choice of the step probabilities.

The pinning transition point in the periodic potential $\beta^{(n)}=\{\beta_1,...,\beta_{n}\}$ with
the period $n$ ($n \gg 1$), where $n$ is assumed to be odd, is determined by the equation
\be
\det \hat T_{s}\{\beta^{(n)}\}\Big|_{s=s_{\rm tr}=1} = 0
\label{eq:47}
\ee
Let us stress once again that we are in the situation where the sequence $\{\beta\}$ consists
of $\ell=N/n$ copies of random subsequences $\{\beta^{(n)}\}$ of length $n$ ($n\gg 1$) each. In
Appendix \ref{sec:appb} we show that the pinning transition point in the chain consisting of $\ell$
($\ell\to\infty$) copies of subsequences $\{\beta^{(n)}\}$ ($n\to\infty$) is the same as in the
single subsequence $\{\beta^{(n)}\}$ (i.e. for $\ell=1$) in the limit $n\to\infty$.

\section{Perturbative calculation of the phase boundary}
\label{sect:4}

In this Section we analyze the spectrum of the matrix $\hat T\equiv \hat T_s\Big|_{s=s_{\rm tr}=1}$
given by Eq.~(\eq{eq:45}) using  standard second order perturbation theory. The disorder is
supposed to be weak, i.e. the fluctuations of the diagonal elements in Eq.~(\ref{eq:47}) are small
compared to their mean value.

\subsection{Non-random substrate}

We begin with the non-random situation and show that in this case Eqs.\eq{eq:45}--\eq{eq:47}
reproduce the results derived in Section~\ref{sect:2:1}. Thus, we shall consider this non-random
case as a reference state and develop a perturbation expansion with respect to this unperturbed
state. In the absence of any disorder Eq.~\eq{eq:47} reduces to
$$
\det \hat T_s(\beta) = \det
\left(\begin{array}{cccc}
    a_0-(\beta-1)^{-1} & a_1 & \dots & a_{n-1} \\
    a_{n-1} & a_0- (\beta-1)^{-1} & \dots  & a_{n-2} \\
    \vdots & \vdots    & \ddots  & \vdots \\
  a_1 & a_2  & \dots & a_0-(\beta-1)^{-1}
\end{array} \right)=0
$$
where the non-random matrix $\hat T_s (\beta)$ is a special case of a Toeplitz matrix known as a
{\em circulant} matrix \cite{grey}. We diagonalize it in the standard way:
\be
\hat T_s= \hat V\hat \Lambda \hat V^\dagger; \qquad\qquad \hat V = \frac{1}{\sqrt{n}}
\left\{e^{-2\pi i m k/n}\,; \quad m,k=0,1,...,n-1 \right\},
\label{eq:55}
\ee
where $\hat \Lambda$ is the diagonal matrix of the eigenvalues
\be
 \lambda_m = \sum_{k=0}^{n-1} a_k e^{-2\pi i m k/n} - (\beta-1)^{-1}  \qquad m=0,1,...,n-1 .
\label{eq:53}
\ee
Using the definition \eqref{eq:46} for the coefficients $a_k$
and by virtue of \eq{eq:53} we have
\be
\lambda_m(s) = \sum_{k=0}^{n-1} \left[\frac{1}{\pi} \int_0^{\pi} \frac{s \sin q\, \sin 2q\,
\cos^{n-1-k} q}{1-s \cos^n q}\,dq \right] e^{-2\pi i m k/n} - (\beta-1)^{-1} .
\label{eq:56}
\ee
At the critical value $s=s_{\rm tr}=1$ the last equation becomes
\be
\lambda_m=2 e^{-4\pi i m/n}-1-2 e^{-4\pi i m/n} \sqrt{1-e^{4 \pi i m/n}}-(\beta-1)^{-1} .
\label{eq:57}
\ee
The transition point $\beta_{\rm tr}$ is determined by the condition $\lambda_0(\beta_{\rm tr})=0$,
yielding the criterion
\be
1-(\beta_{\rm tr}-1)^{-1} = 0; \qquad \beta_{\rm tr} =  2
\label{eq:58}
\ee
which reproduces the solution of the non-random problem
considered at length in Section \ref{sect:2:1}.

\subsection{Random substrate}

Suppose now that the energies $u_m=\ln \beta_m$ for $m=1,...,n$ are independent random variables
distributed according to the law
\be
u_m = w+\vep\sigma_m \qquad\qquad
\sigma_m =
\begin{cases} +1 & \mbox{with the probability $\frac{1}{2}$} \\
-1 & \mbox{with the probability $\frac{1}{2}$} \end{cases}
\label{eq:48}
\ee
and find eigenvalues and eigenvectors of the perturbed matrix $T$ in the second order of series
expansion in the strength of the disorder. Expanding the diagonal elements of the matrix $\hat T$
(Eq.\eq{eq:45}) in $\vep\ll 1$, we get
\be
Q_i=a_0-(e^{u_i}-1)^{-1} = a_0-\left(e^{w+\sigma_i \vep}-1\right)^{-1}; \qquad (i=1,2,...,n)
\label{eq:59}
\ee
Expanding \eq{eq:59}, we arrive finally at the following expression for the diagonal elements $Q_i$
of the matrix $\hat T$
\be
Q_i=a_0-(e^w -1)-p_0+p_1 \sigma_i
\label{eq:60}
\ee
where
\be
\begin{cases} \disp p_0=\frac{1}{2}\frac{e^w(e^w+1)}{(e^w-1)^3}\,\vep^2 \medskip \\
\disp p_1=\frac{e^w}{(e^w-1)^2}\,\vep
\end{cases}
\label{eq:61}
\ee
The matrix $\hat T_s \{\beta^{(n)}\}$ up to the second order perturbation on $\vep$ reads
\be
\hat T_s \{\beta^{(n)}\} = \hat A +\hat B = \hat T_s (e^w) -p_0 \hat I + \left(\begin{array}{cccc}
    p_1 \sigma_1 & 0 &  \dots & 0 \\
    0 & p_1 \sigma_2 &        & 0 \\
    \vdots &  &  \ddots  & \vdots \\
     0 & 0 &  \dots &  p_1 \sigma_n
  \end{array} \right)
\label{eq:62}
\ee
where the matrices $\hat A=\hat T_s (e^w) -p_0  \hat I$ and $\hat B$ are correspondingly the
unperturbed and perturbed parts of the matrix $T$ (by $\hat I$ we have denoted the unit $n \times
n$ matrix).

Standard 2nd order perturbation theory leads to the following expression for the perturbed
eigenvalue $\lambda_m'$:
\be
\lambda_m'\{\beta^{(n)}\}=\lambda_m + {\bf v}_m \hat B\; {\bf v}^{\top}_m + \sum_{m'\neq m, \nu}
\frac{\left|{\bf v}^{(\nu)}_m \hat B\; {\bf v}^{\top,(\nu)}_{m'}\right|^2}{\lambda_m-\lambda_{m'}}
\label{eq:63}
\ee
where $\nu$ is the degeneracy of the eigenvalues and the eigenvectors ${\bf v}_m $ are the columns of the
matrix \eqref{eq:55}.  In what follows we are interested in the
computation of the largest eigenvalue $\lambda_0'$.

For any particular sequence $\{\beta^{(n)}\}$ the transition point is determined by the condition
\be
0=\lambda'_0\{\beta^{(n)}\}=1-\frac{1}{e^w-1}-p_0+\frac{p_1}{n}\left(\sum_{k=0}^{n-1}\sigma_k\right)
+ \frac{p_1^2}{n^2}\left(\sum_{m'\neq 0} \frac{1}{\lambda_0-\lambda_{m'}}\sum_{k=0}^{n-1}
\sum_{k=0}^{n-1}\sigma_k\sigma_{k'}e^{-2\pi m'(k-k')/n}\right)
\label{eq:transition2}
\ee
Let us denote
\be
\left\{\begin{array}{l}
\disp S_1=\sum_{k=0}^{n-1}\sigma_k \\
\disp S_2=\sum_{m'\neq 0} \frac{1}{\lambda_0-\lambda_{m'}}\sum_{k=0}^{n-1}
\sum_{k=0}^{n-1}\sigma_k\sigma_{k'}e^{-2\pi m'(k-k')/n}
\end{array} \right.
\label{eq:i1}
\ee
and expand the coefficients $p_0$ and $p_1$ for small $w$ in the vicinity of $\ln 2$ up to the 2nd
order in $w$ and $\vep$:
\be
\begin{array}{l}
\disp p_0=3\vep^2+13\vep^2(\ln 2-w) + \frac{75}{2} \vep^2(\ln 2-w)^2 \medskip \\
\disp p_1=2\vep+6\vep(\ln 2 - w)+13\vep (\ln 2-w)^2
\end{array}
\label{eq:coeff}
\ee
In the 2nd order perturbation series we keep only the terms up to the second order in $\tilde w=\ln
2 - w$ and $\vep$:
\be
\lambda'_0=-2\tilde w - 3 {\tilde w}^2-3\vep^2 + (2\vep + 6 \vep \tilde w) \frac{S_1}{n}+ 4\vep^2
\frac{S_2}{n^2} = 0
\label{eq:transition3}
\ee
Solving the quadratic equation $\tilde w(\vep)=0$ (we have chosen the branch, on which $w\to \ln 2$
as $\vep\to 0$) and re-expanding the final expression in $\vep$ up to the 2nd order, we get
\be
\tilde w = - \vep \frac{S_1}{n} + \frac{1}{2} \vep^2 \left(3 \frac{S_1^2}{n^2} + 4
\frac{S_2}{n^2}-3\right)
\label{eq:transition4}
\ee
Let us stress that the expression \eq{eq:transition4} is valid for any distribution ${\cal
P}\{\sigma\}$ of random Ising--type variable $\sigma_i$. The general form of the transition curve
in the 2nd order approximation reads
\be
\tilde w = - \vep \frac{\la S_1 \ra}{n} + \frac{1}{2} \vep^2 \left(3 \frac{\la S_1^2 \ra}{n^2}+ 4
\frac{\la S_2 \ra}{n^2}-3\right)
\label{eq:transition5}
\ee
The terms proportional to $\la S_1\ra$  vanish due to the symmetry $\sigma_i \leftrightarrow -\sigma_i$. The
calculation of the second order correction is straightforward though rather tedious and is given in
Appendix \ref{sec:appa}. The final result for the averaged transition point $w$ in the limit of
large $n$ reads
\begin{equation}
\tilde w = \frac{\vep^2}{2}
\label{eq:transition6}
\end{equation}
so that the phase diagram is determined by the equation
\be
w=\ln 2 - \frac{\vep^2}{2}
\label{eq:72}
\ee
This result of the  perturbation theory for the phase boundary   $w(\vep)$ is shown in \fig{fig:phase}
by dashed line. For  $\vep < 0.5$ there is a good agreement of the perturbative approach with the numerical
data and results of our renormalization group computations of the next Section.

\section{Simple RG consideration}
\label{sect:rg}

The location of the localization (wetting) transition can be found following a simple
renormalization argument. We start with the master equation (\ref{eq:7}) in the Fourier space
\begin{eqnarray}
  \label{eq:7_fourrier}
  G_{N+1}(q) = \cos q\, G_N (q) +\big(\beta_N - 1\big) \frac{\sin q}{\pi}\int_0^\pi
  \!dq'\, \sin 2q' \,G_N(q')
\end{eqnarray}
and iterate it twice
\begin{eqnarray}
  \label{eq:7_fourrier_twice}
  G_{N+2}(q) &=& \cos^2 q\, G_N (q) \\
  &+&\big(\beta_N - 1\big) \frac{\sin q\cos q}{\pi}\int_0^\pi
  \!dq'\, \sin 2q'\, G_N(q')+\big(\beta_{N+1} - 1\big) \frac{\sin q}{\pi}\int_0^\pi
  \!dq'\, \sin 2q'\cos q'\, G_N(q') \\
  &+&\big(\beta_N - 1\big) \big(\beta_{N+1}-1\big) \frac{\sin q}{\pi^2}\int_0^\pi
  \!dq'\, \sin 2q' \sin q' \int_0^\pi \!dq''\, G_N(q'').
\end{eqnarray}
The last term in the r.h.s. vanishes due to the orthogonality. The second and third term describe
processes of two--step arrival to $x=1$. We note that in the long--wavelength limit $q\to 0$ they
are identical as seen by  replacing  the cosines by unity. This corresponds to neglecting variation
in $G_N (x)$ on  scale of the lattice spacing and will be justified later. Combining these terms we
obtain an expression similar to the small $q$ version of Eq.(\ref{eq:7_fourrier}) with modified
random part
\begin{eqnarray}
  \label{eq:mod_rand_part}
  G_{N+2}(q) - G_N (q)=  -q^2  G_N (q) +\big(\beta_N+\beta_{N+1} - 2\big) \frac{\sin q}{\pi}\int_0^\pi
 \!dq'\, \sin 2q' G_N(q')
\end{eqnarray}
and corresponding to the double time interval. By this procedure the random potential has been
normalized to the arithmetic mean of two  subsequent terms:
\begin{eqnarray}
  \label{eq:rand_ren}
  \beta'_{N/2} = \frac{\beta_N+\beta_{N+1}}{2}
\end{eqnarray}

The probability distribution of the renormalized disorder $\beta'$ is related to that of $\beta$ as
follows. Let $P(\beta)$ be a probability distribution of $\beta_j$ and $P'(\beta')$ the one of the
new variables. The corresponding Fourier transforms (characteristic functions) $\xi (\lambda)$ and
$\xi' (\lambda')$ are related as
\begin{eqnarray}
  \label{eq:rg_proba}
  \xi'(\lambda') = \xi^2 (\lambda'/2).
\end{eqnarray}
The fixed point of this transformation is the exponential function
\begin{eqnarray}
  \label{eq:sol_rg}
  \xi'(\lambda) = \xi(\lambda) =\exp(i\lambda\la\beta\ra)
\end{eqnarray}
which is just a consequence of the Central Limit Theorem.

\begin{figure}[ht]
\epsfig{file=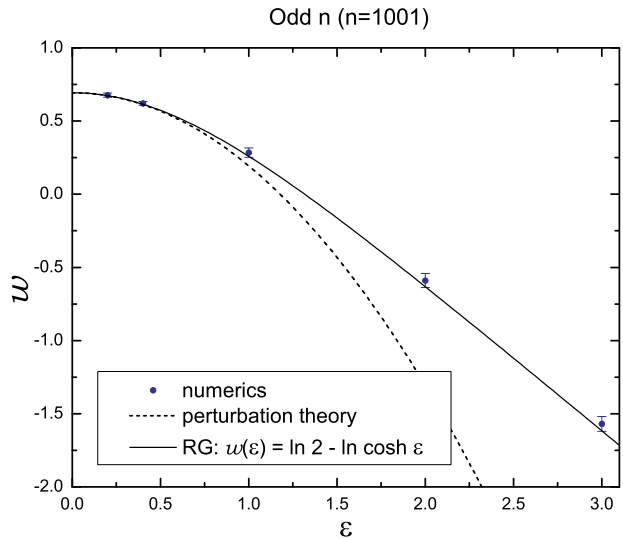, width=10cm} \caption{Phase diagram for quenched disorder in the plane
$(\vep,w)$. The renormalization group computations and perturbation theory are depicted by solid
and dashed lines correspondingly. The results of numerical averaging are shown for odd $n$
($n=1001$).}
\label{fig:phase}
\end{figure}

Therefore the disorder  $\beta_j=e^{u_j}$ can be just replaced by their mean value
$\left<\beta_j\right>=\left<e^{u_j}\right>$ in the long--wavelength limit. The crucial  observation
is that $\left<e^{u_j}\right>\neq e^{-\left<u\right>}$, so fluctuations of $u_j$ modify the
localization criterion. For disorder discussed in the previous section (see \eq{eq:48}). The mean
value of $\beta_j$ is given by
\begin{eqnarray}
  \label{eq:beta_mean}
  \left<\beta_j\right> = e^{w}\cosh \vep
\end{eqnarray}
so it depends on both parameters, the mean value, $w$, and fluctuations, $\vep$. The localization
criterion reads from (\ref{eq:20})
\begin{eqnarray}
  \label{eq:loc_crit}
  \bar{\beta} = e^{w}\cosh\epsilon = 2 \qquad \Rightarrow \qquad w=\ln 2 -\ln\cosh\vep
\end{eqnarray}
The lowest order in expansion of this expression in powers of $\vep$ yields the  result of the
perturbation theory \eqref{eq:72}. The graphical representation of the RG phase boundary
\eq{eq:loc_crit} is shown in Fig.\ref{fig:phase} by solid line.

Finally, we observe that from the expressions (\ref{eq:13})--(\ref{eq:14}) it follows that the
bound state contribution to the Fourier transform
\begin{eqnarray}
  \label{eq:GNq}
  G_N(q) \sim \frac{\beta-2}{2\beta}\frac{\sin q}{1-s_0 \cos q} s_0^{-N}  ,
  \qquad\qquad s_0=\frac{2\sqrt{\beta-1}}{\beta}
\end{eqnarray}
is strongly peaked around $q=0$ close to the transition point $\beta=2$. This justifies replacing
$\cos q$ by unity in the integrals in Eq.(\ref{eq:7_fourrier_twice}).

\section{Numerical analysis of the depinning transition}
\label{sect:3}

\begin{figure}[ht]
\epsfig{file=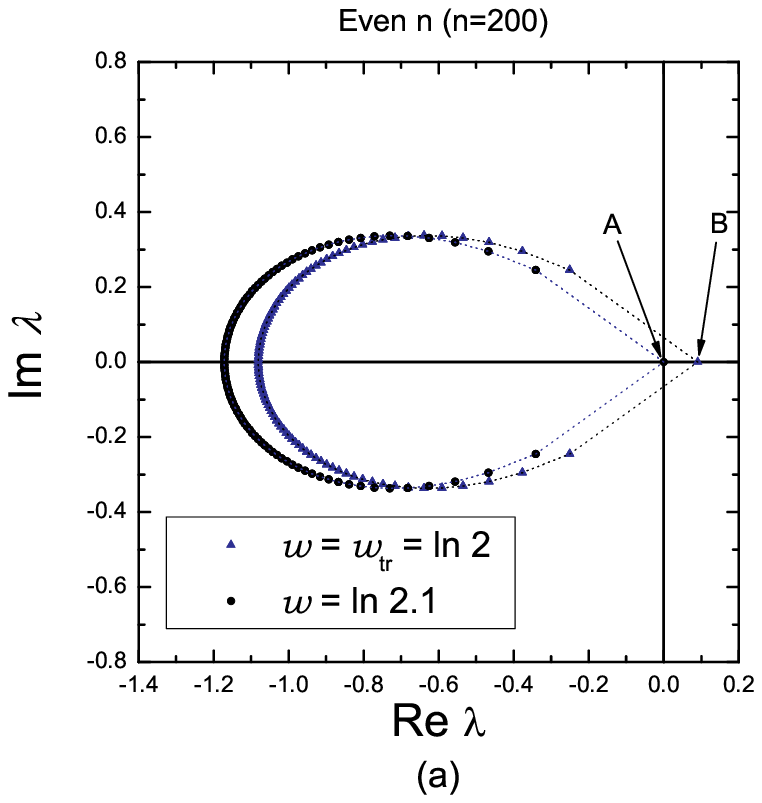, width=8cm} \epsfig{file=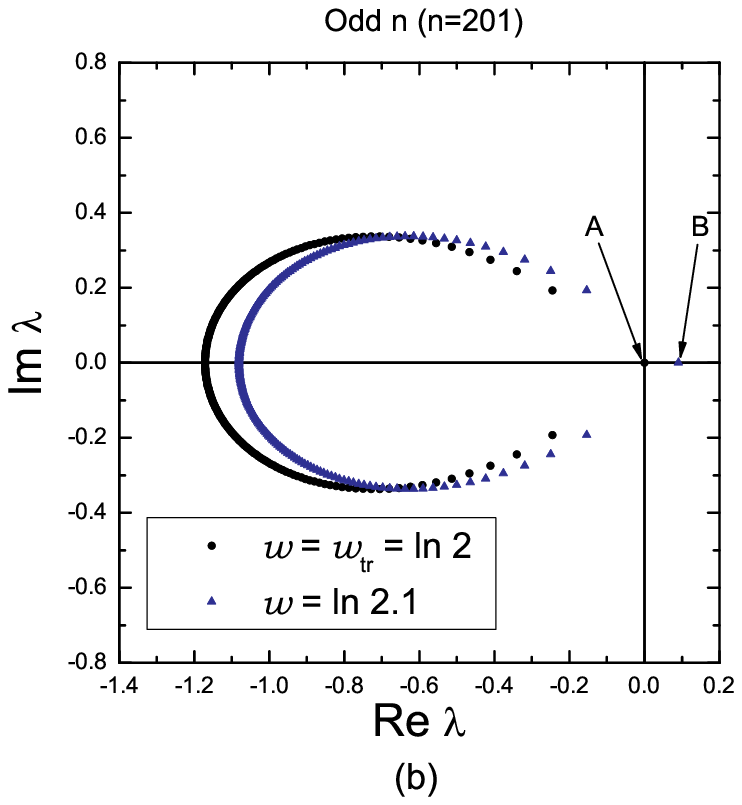,width=8cm} \caption{The
eigenvalues $\lambda_m$ of the matrix $\hat{T}$ in the complex plane $({\rm Re} \lambda, {\rm Im}
\lambda)$ for: (a) even $n$ ($n=200$) and (b) odd $n$ ($n=201$) for two different values of $w$:
$w= \ln 2.0$ (transition point) and $w = \ln 2.1$.} \label{fig:zeros}
\end{figure}

To visualize the eigenvalues of the matrix $\hat T$ , consider the case of absence of any disorder,
i.e. take in Eq.\eq{eq:45} the homogeneous sequence $\{\beta^{(n)}\}$ with $\beta_j\equiv \beta$ for
all $j=1,...,n$. In \fig{fig:zeros} we have plotted the eigenvalues $\lambda_m$ ($m=1,...,n$) of
the matrix $\hat T$ in the complex plane for two values $w=\ln 2.1$ and $w=\ln 2.0$ (transition point)
for even (a) and odd (b) values of $n$. For even values of $n$ the spectrum of the
pure matrix $\hat T$ is doubly degenerate, while for odd $n$ this degeneracy
disappears. In what follows we consider odd $n$ only.

We now consider disorder generated by the distribution (\ref{eq:48}) and find numerically the
location of the pinning transition. The general procedure is very simple: we fix some value $\vep$,
then generate ensemble of random sequences $\beta^{(n)}$ from the distribution \eqref{eq:48}
and find for each sequence such a $w$ at which the  real
eigenvalue of the matrix $\hat T$ crosses zero implying the condition
\be
\lim_{n\to\infty}\det \hat T\{\beta_1,...,\beta_n\} = 0
\label{eq:det}
\ee
The obtained critical value is then averaged over realizations $\{\beta^{(n)}\}$ of quenched disorder.
We take the size of the $n\times n$ matrix $\hat{T}$ equal to $n=1001$.  In
\fig{fig:phase} we show  the points corresponding to the averaged phase boundary for the ensemble
of quenched sequences $\{\beta^{(n)}\}$ in the  space of the parameters $\vep,w$.
\begin{figure}[ht]
\epsfig{file=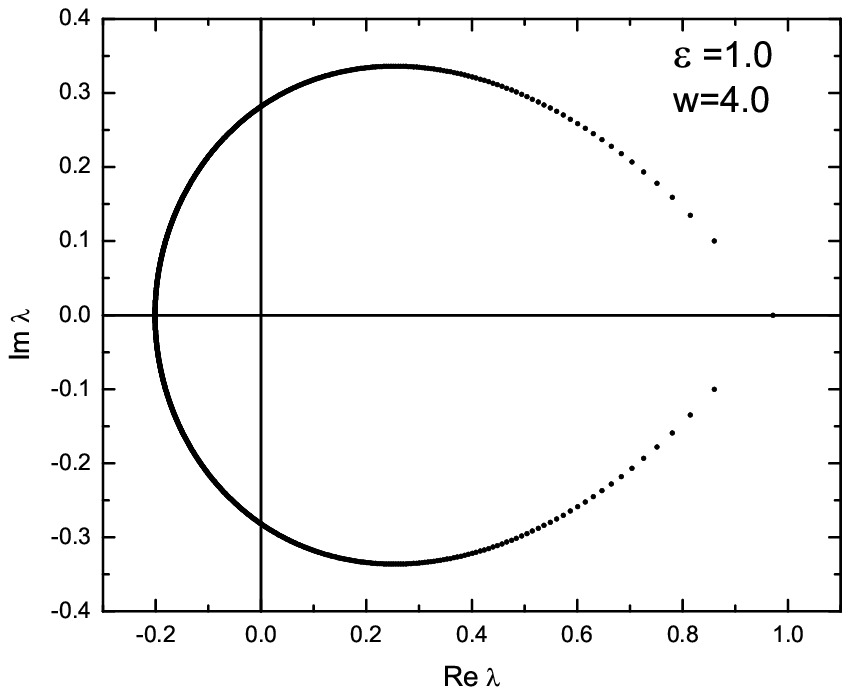, width=4cm} \epsfig{file=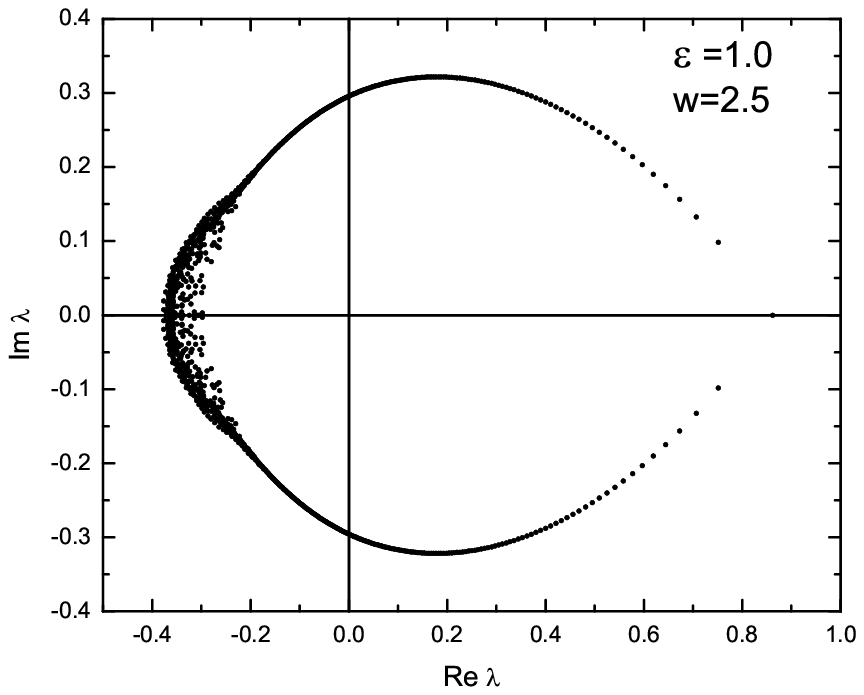, width=4cm} \epsfig{file=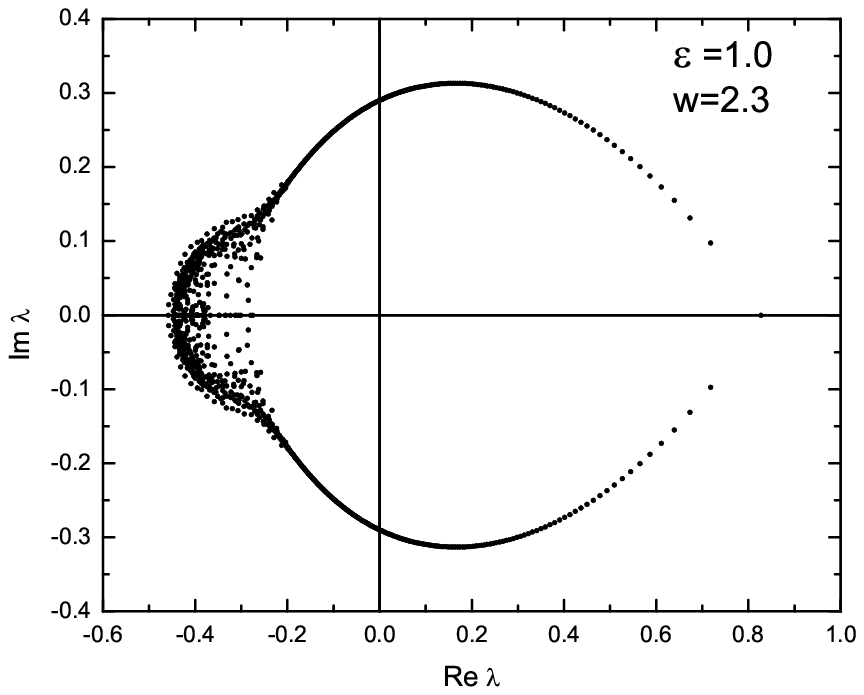,
width=4cm} \epsfig{file=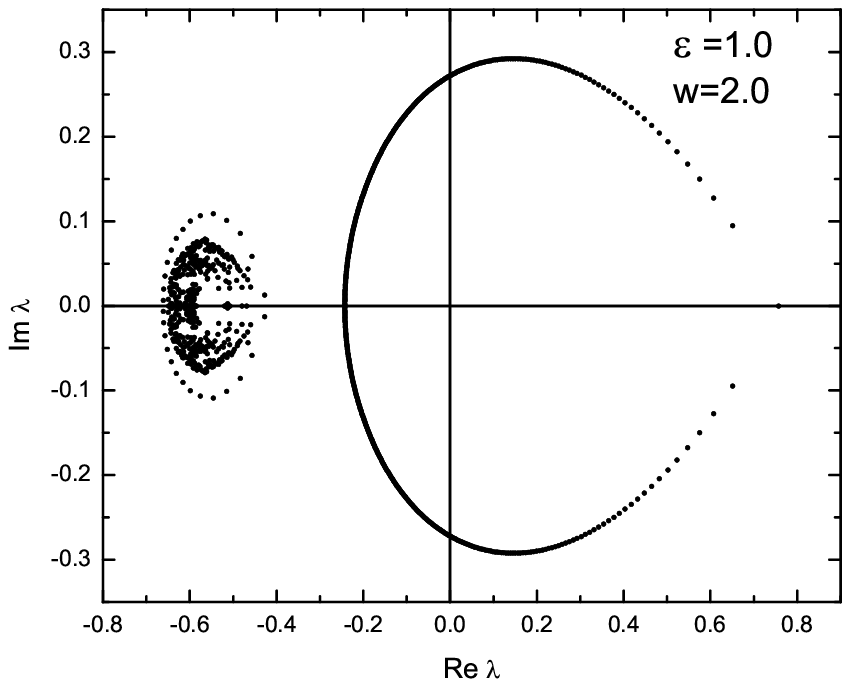, width=4cm} \\ \epsfig{file=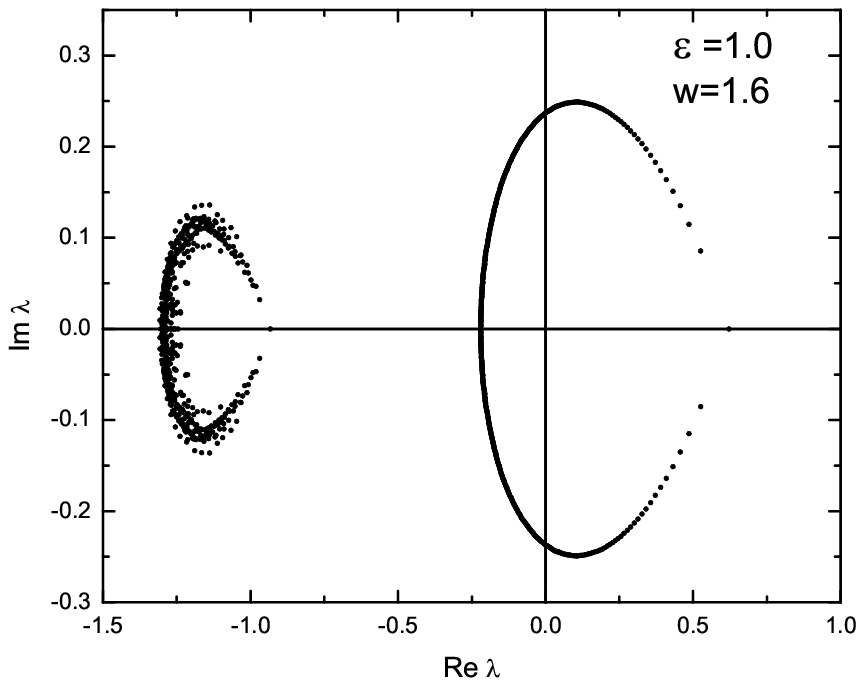, width=4cm}
\epsfig{file=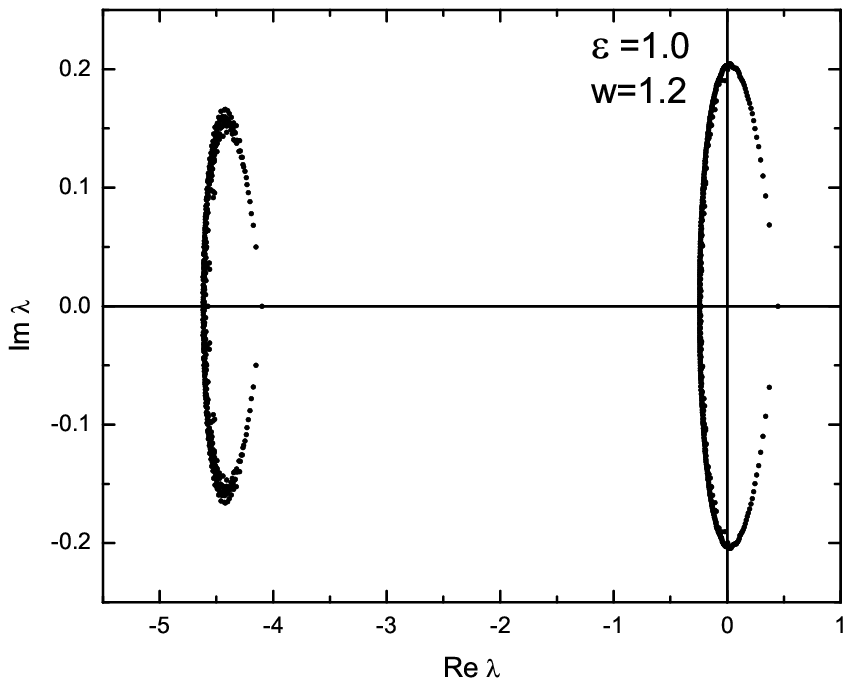, width=4cm} \epsfig{file=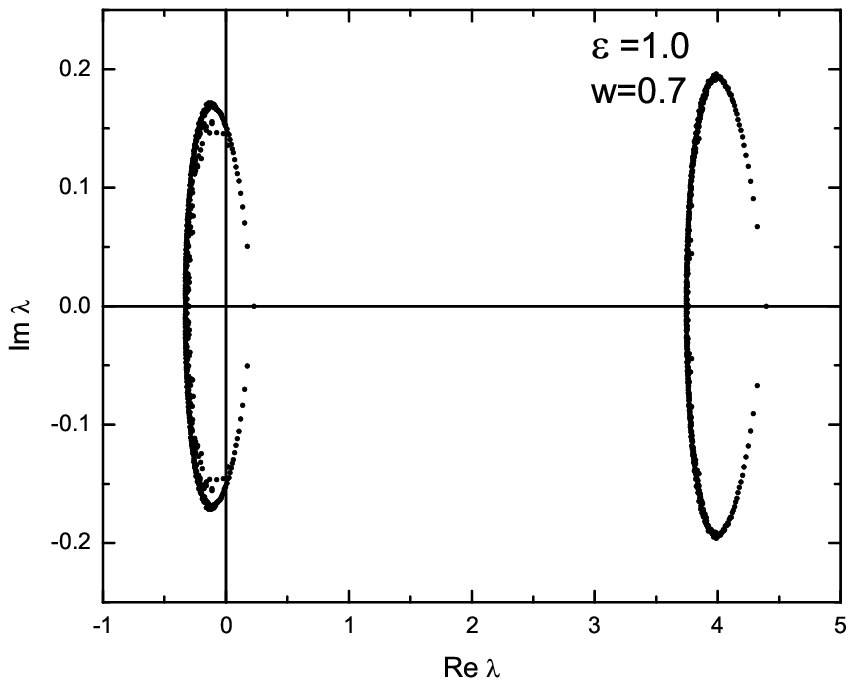, width=4cm} \epsfig{file=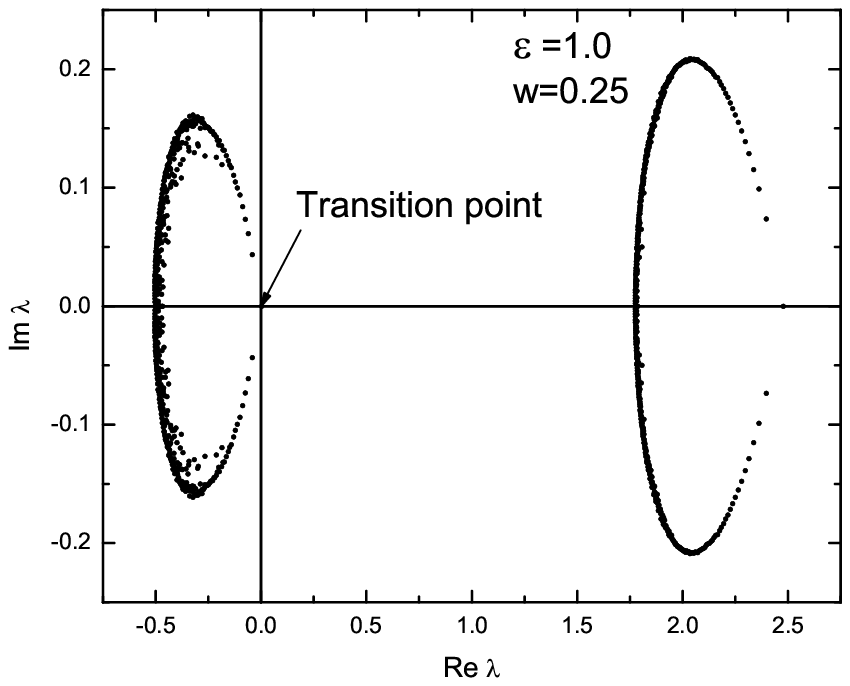,
width=4cm} \\ \epsfig{file=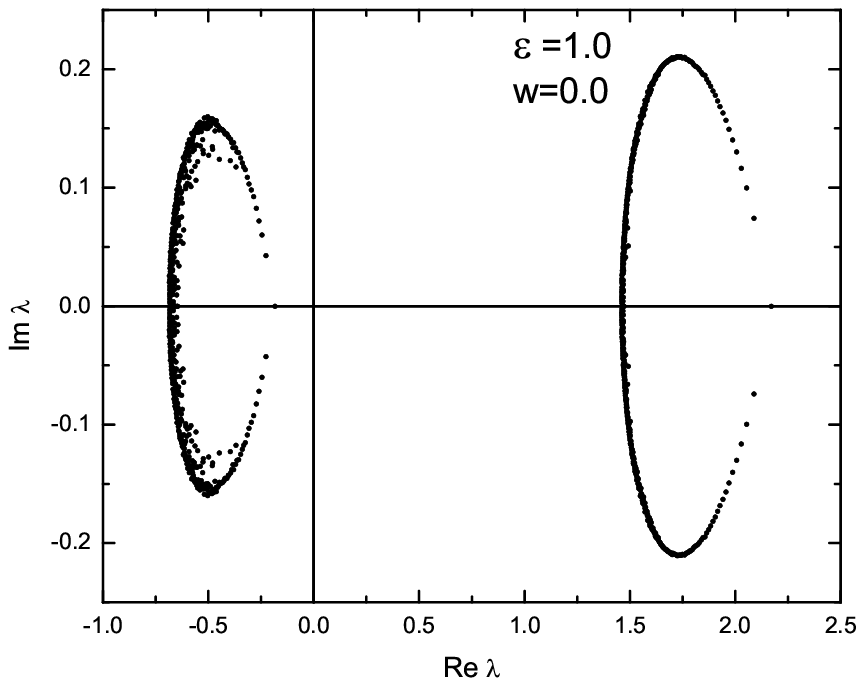, width=4cm} \epsfig{file=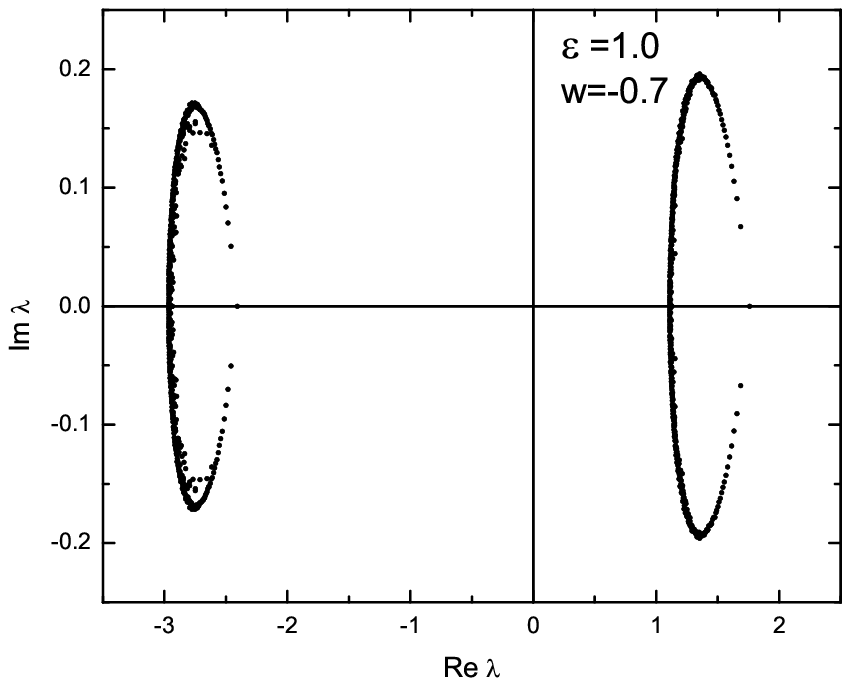, width=4cm}
\epsfig{file=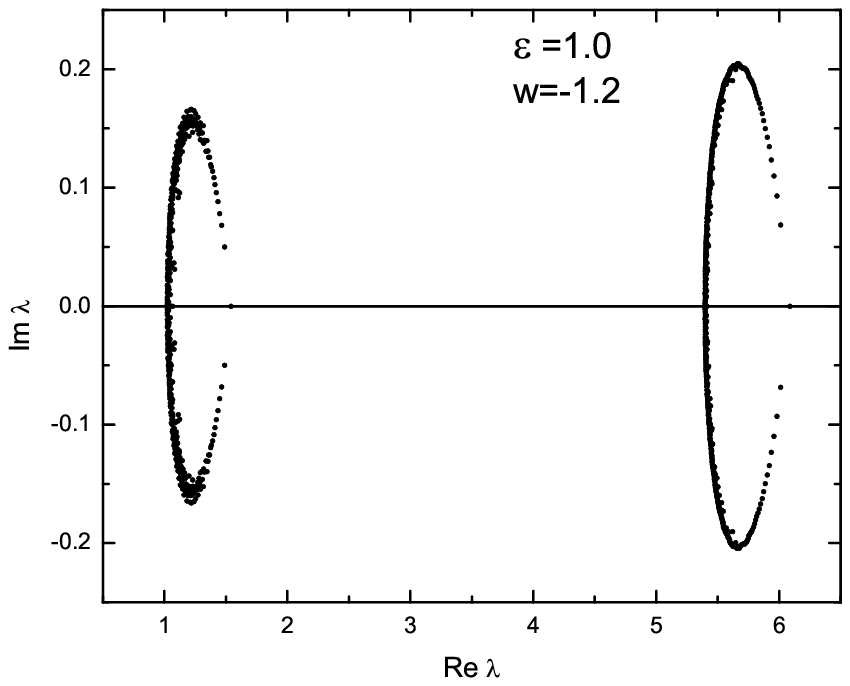, width=4cm} \epsfig{file=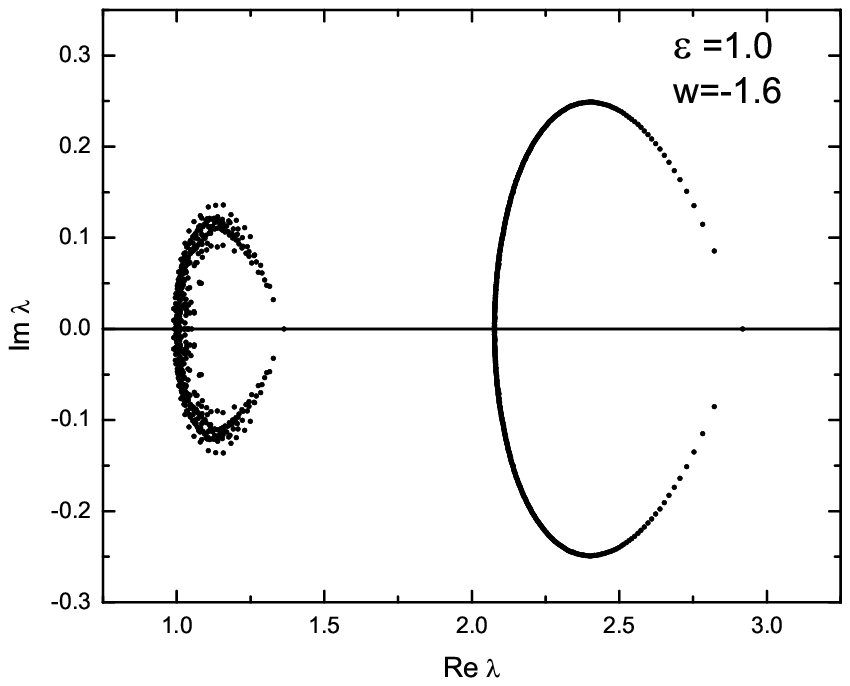, width=4cm} \\ \epsfig{file=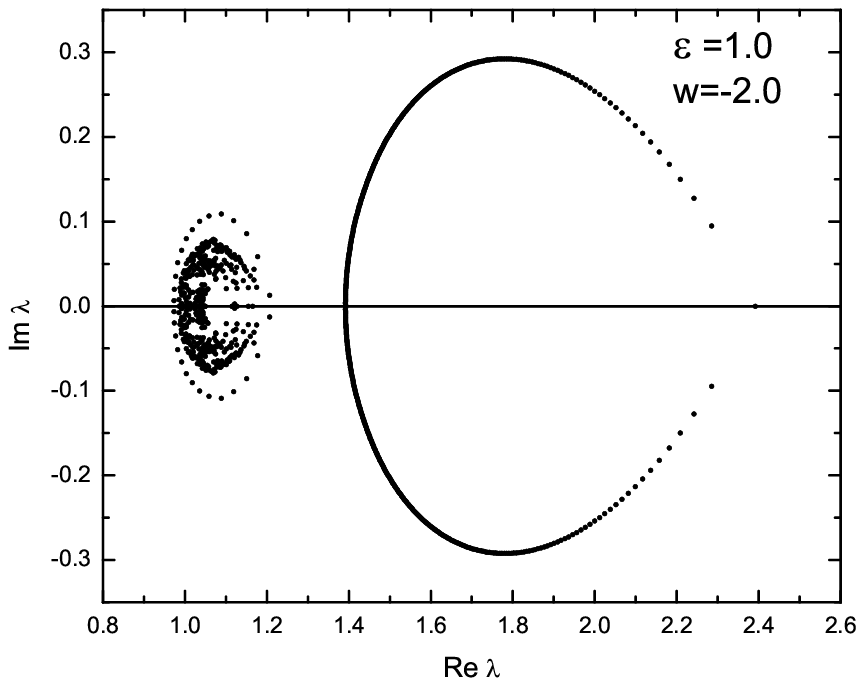,
width=4cm} \epsfig{file=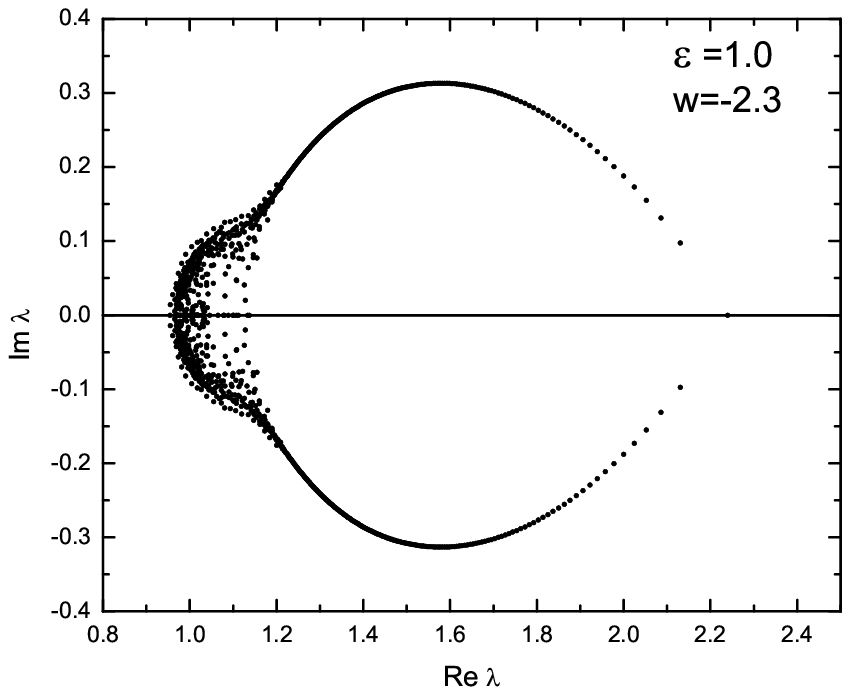, width=4cm} \epsfig{file=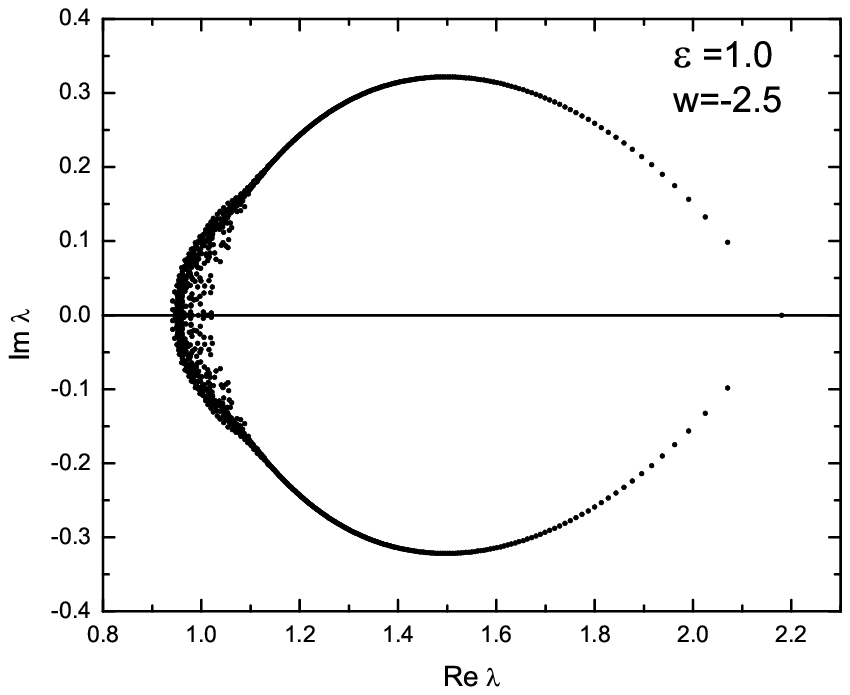, width=4cm}
\epsfig{file=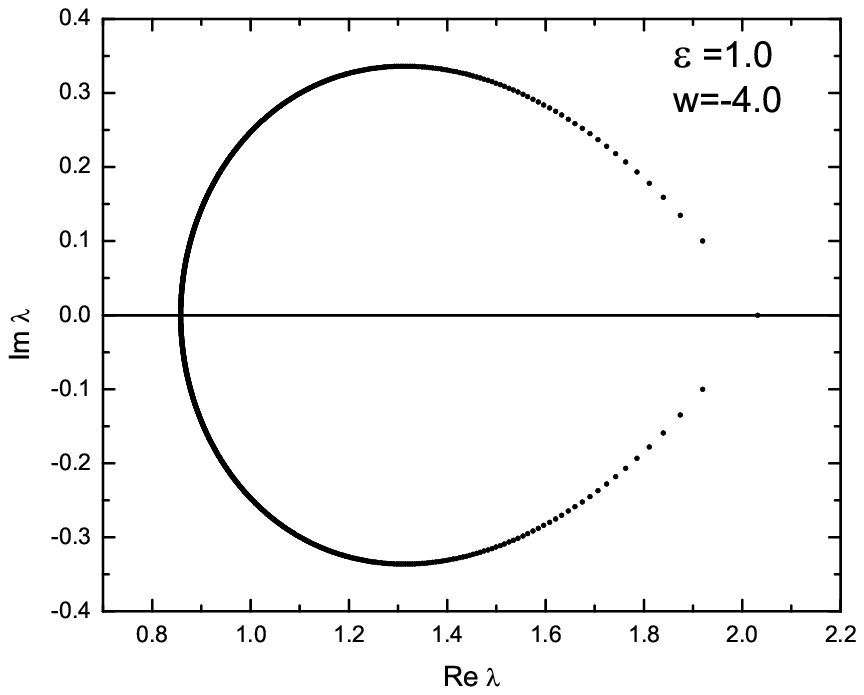, width=4cm} \caption{Sample plots of the location of eigenvalues of the
matrix $\hat T$ for $\vep=1.0$ and different values of $w$ ($n=1001$).}
\label{fig:sample}
\end{figure}
In the next section we discuss the obtained numerical phase diagram.

\section{Conclusions}
\label{sect:concl}

In this paper we have discussed the problem of wetting of a one--dimensional random walk on a
half--line $x\ge 0$ in a short--ranged periodic potential located at the origin $x=0$. When the
period of the potential becomes equal to the length of the random walk trajectory, then one could
conjecture that the trajectory under consideration is interacting with the quenched random
potential. In other words, we approach the random potential, increasing more and more the period of
the substrate potential.

The most attention is paid to the question concerning the location of the transition point averaged
over all equally weighted realizations of the disorder. We have shown that for any specific
disorder realization, the transition point can be obtained from the condition that the determinant
of the matrix \eq{eq:45} (with the entries \eq{eq:46}) is equal to zero at $s=1$ --- see
Eq.\eq{eq:47}. The period of the substrate is given the size $n$ of the matrix \eq{eq:45} which
remains to be very large but finite in our consideration, and the fugacity $s$ is conjugate to the
number of copies, $\ell=N/n$, of $n$--periodic potential (so, $N$ is the total length of the
substrate). Setting the fugacity $s$ in \eq{eq:45}--\eq{eq:46} to its limiting value $s_{\rm
tr}\equiv 1$ amounts to the consideration of $\ell\to \infty$ copies of the system.

One may worry that in the thermodynamic limit the transition point in the periodic sequence of
$\ell$ ($\ell\to\infty$) subchains each of length $n$ ($n\gg 1$) differs from the transition point
in a single ($\ell=1$) subchain (of the same primary structure) of length $n$ ($n\to\infty$).
However this is not true and in Appendix \ref{sec:appb} we show that the transition point is not
sensitive to the sequence of thermodynamic limits and for $n\to\infty$ is independent on number of
copies, $\ell$.

We have compared the results of our numerical simulations performed in a wide range of disorder
strengths with the results of the perturbation theory for weak disorder, and the results of a
simple renormalization group consideration. The results obtained allow us to conclude that for weak
disorder $\vep <0.5$  all approaches (including the perturbation theory) are in a good agreement. For
sufficiently high disorder strengths $\vep>\vep_{\rm pert}\sim 0.5$ the second order
perturbation theory fails, while the results of the renormalization group always agree with the
numerical simulations --- see the \fig{fig:phase}. This point deserves some special
discussion. Namely, recall that results of the RG computations claim the existence of the averaged
Boltzmann weight which governs the behavior of the system under renormalization. This fact signals
the existence of an effective preaveraged (with respect to the disorder) annealed system with
averaged Boltzmann weight $\left<e^{u_m}\right>$. This statement is consistent with the conclusion of
works \cite{forgacs,gro}, but contradicts  the statement that the disorder is marginally
relevant formulated for the first time in \cite{derrida}.

Let us note at the very end that the advantage of our matrix approach is its transparency.
We have made no any uncontrolled assumption and have reduced very complicated initial problem with
many degrees of freedom to the well posed problem of finding eigenvalues of some random matrix with
a relatively simple structure. The localization criteria $\det T\{\beta_1,...,\beta_N\}=0$ for a
random sequence $\{\beta_1,...,\beta_N\}$ in the limit $N\to\infty$ is {\em exact} for any
strengths and any distributions of $\beta_i$ ($i=1,...,N$).

The results obtained in our work allow us to state that even for strong disorder the shift of the
averaged pinning point of the random walk in the ensemble of random realizations of substrate
disorder {\em is indistinguishable} from the pinning point of the system with preaveraged (i.e.
annealed) Boltzmann weight.

\acknowledgments

The authors would like to thank Martin Long for illuminating discussions at the final stage of this
work. S.N. is grateful to Alexey Naidenov for valuable discussions of random matrix approach in
course of collaboration on related problem \cite{nech_nai} and to Gambattista Giacomin for useful
coments. The work is partially supported by the grant ACI-NIM-2004-243 "Nouvelles Interfaces des
Math\'ematiques" (France), by the National Science Foundation under Grant No. PHY05-51164 and by
the EPSRC Advanced Fellowship EP/D072514/1.

\begin{appendix}

\section{Perturbation theory for matrix $T_s$}
\label{sec:appa}

The terms in Eq. \eq{eq:transition5} are
\be
\left\{\begin{array}{l}
\disp \frac{1}{n}\la S_1 \ra=\frac{1}{n}\sum_{k=0}^{n-1}\la \sigma_k \ra \\
\disp \frac{1}{n^2}\la S_1^2 \ra=\frac{1}{n^2}\sum_{k=0}^{n-1}\sum_{k'=0}^{n-1}\la \sigma_k
\sigma_{k'}\ra \\ \disp \frac{1}{n^2} \la S_2 \ra =\frac{1}{n^2} \sum_{m'\neq 0}
\frac{1}{\lambda_0-\lambda_{m'}} \sum_{k=0}^{n-1} \sum_{k=0}^{n-1}\la \sigma_k\sigma_{k'} \ra
e^{-2\pi m'(k-k')/n}
\end{array} \right.
\label{eq:65}
\ee

The computation of $\la S_1 \ra$ and $\la S_1^2 \ra$ where averaging $\la ...\ra$ is performed over
the symmetric distribution ${\cal P}\{\sigma\}$, is straightforward:
\be
\frac{1}{n}\la S_1 \ra=0; \quad \frac{1}{n^2}\la S_1^2 \ra=\frac{1}{n}
\label{eq:65a}
\ee

Using \eq{eq:57}, we obtain for the unperturbed part $\hat A$ of the matrix $\hat T$
(Eq.\eq{eq:62}) the following expression for the eigenvalues
\be
\lambda_m=2 e^{-4\pi i m/n}-1-2 e^{-4\pi i m/n} \sqrt{1-e^{4 \pi i m/n}}-(e^w-1)^{-1}-p_0
\label{eq:68}
\ee
Taking into account that
$$
\sqrt{1-e^{4 \pi i m/n}}=\sqrt{2 \sin\frac{2\pi m}{n}}\; e^{i \pi (m/n-1/4)},
$$
we can substitute \eq{eq:68} into last line of \eq{eq:65}. Thus we get
\be
\frac{1}{n^2}\la S_2 \ra = \frac{1}{n} \sum_{m'\neq 0} \frac{1}{\lambda_0-\lambda_{m'}} =
\frac{1}{n} \sum_{m'\neq 0}^{n-1} \left( 2-2e^{-4\pi i m'/n}+ 2 e^{-\pi i(1/4+3m'/n)}
\sqrt{2\sin\frac{2\pi m'}{n} }\right)^{-1}
\label{eq:69}
\ee
Since the eigenvalues are distributed symmetrically with respect to the real axis (see
\fig{fig:zeros}a), the sum $\sum_m \frac{1}{\lambda_0-\lambda_{m'}}$ takes only the real values.
After some algebra, we arrive at the final equation for the expectation $\la S_2\ra$:
\be
\frac{1}{n^2}\la S_2\ra = \frac{2}{n} \sum_{m'\neq 0}^{(n-1)/2}\; \frac{1-\cos\frac{4\pi m'}{n} +
\sqrt{2 \sin \frac{2 \pi m'}{n}} \cos\left(\frac{3\pi m'}{n} + \frac{\pi}{4} \right)}
{\sin\frac{2\pi m'}{n}\left(1-2\sqrt{2 \sin \frac{2 \pi m'}{n}} \sin\left(\frac{3\pi m'}{n} +
\frac{\pi}{4}\right) + 2\sin\frac{2\pi m'}{n} \right)}
\label{eq:70}
\ee
In the limit $n\to\infty$ the sum in \eq{eq:70} can be replaced by the integral:
\be
\frac{1}{n^2}\la S_2\ra  \approx 2 \int_0^{1/2} \frac{1-\cos 4\pi x + \sqrt{2 \sin 2\pi x}
\cos\left(3\pi x + \frac{\pi}{4}\right) } {\sin 2\pi x \left(1-2\sqrt{2 \sin 2\pi x} \sin
\left(3\pi x + \frac{\pi}{4}\right) + 2\sin 2\pi x \right) } dx = 1
\label{eq:71}
\ee
Substituting \eq{eq:65a} and \eq{eq:71} into \eq{eq:transition5} we arrive in the limit
$n\to\infty$ at the desired equation \eq{eq:transition6}.

\section{Independence of phase boundary on number of periods $\ell$ in the thermodynamic limit}
\label{sec:appb}

It follows from the general definition of the generating functions, that in order to consider the
phase transition in an individual period of the substrate potential \eq{eq:periodic}, i.e. for
$\ell=1$, we should shift the fugacity $s$ (conjugated to $\ell$) away from its marginal value
$s=s_{\rm tr}=1$ (corresponding to $\ell\to\infty$) to some value $s_{\ell}$ lying in the interval
$0<s_{\ell}<s_{\rm tr}$.

The number of periods, $\ell$, is controlled by the fugacity $s_{\ell}$. So, to compare the
transition points in chains of different periods, we should normalize the Boltzmann weights of
chain of $\ell$ periods per a weight of a a single period, $s_{\ell}$ (for $\ell=1$). Dividing the
Boltzmann weights in the matrix $T_s$ (see Eq.\eq{eq:45}) by $s_{\ell}$, we get: $(\beta-1) \to
(\beta^*-1)/s_{\ell}$. Now we could investigate how the normalized transition point, $\beta^*_{\rm
tr}$, depends on the typical fugacity $s_{\ell}$. It is easy to see that in the absence of any
disorder (i.e. for $\vep=0$) the eigenvalue $\lambda_0$ (see Eq.\eq{eq:56} for $m=0$) is given by
the following integral
\be
\lambda_0(s_{\ell}) = \frac{s_\ell}{\pi} \int_0^{\pi} \frac{\sin q\, \sin 2q\, (1-\cos^n
q)}{(1-s_{\ell}\, \cos^n q)(1-\cos q)}\,dq - \frac{s_{\ell}}{\beta^*-1}
\label{eq:56new}
\ee
The equation $\lambda_0(s_{\ell})=0$ in the limit $n\to\infty$ determines the position of the
normalised transition point, $\beta^*_{\rm tr}(s_{\ell})$. One can easily verify that in the limit
$n\to\infty$ the integral \eq{eq:56new} is independent of $s_{\ell}$ and, hence, $\beta^*_{\rm
tr}(s_{\ell})=2$ for any $0<s_{\ell}<s_{\rm tr}$. The same conclusion holds for $\vep\neq 0$ in the
matrix $T_s$: the transition point in the limit $n\to\infty$ for any random primary sequence is
independent of the effective fugacity $s_{\ell}$ corresponding to finite number of periods, $\ell$.

So, we conclude that in the thermodynamic limit the transition point in the periodic sequence of
$\ell$ ($\ell\to\infty$) random subchains each of length $n$ ($n\gg 1$) coincides with the
transition point in a single ($\ell=1$) random subchain (of the same primary structure) of length
$n$ ($n\to\infty$).

This conclusion has very transparent physical sense. Namely, consider from the very beginning the
random walk with {\em periodic boundary conditions} in the space. The simplest way is to suppose
that the first ($t=0$) and last ($t=N$) steps of the random trajectory are always attached to the
random substrate at the point $x=0$. For such a periodic system the location of the transition
point is exactly given by the equation
$$
\lim_{N\to\infty}\det T\{\beta_1,\beta_2,...,\beta_N\}=0
$$
Hence, the transition point is not sensitive to whether the terminal step of the random walk is
attached to the surface or not. This is evident in the localized regime where the fluctuations of
mean--square end-to-end distance are constant (see Eq.\eq{eq:7}).

\end{appendix}

\end{document}